# Secure two-way fiber-optic time transfer against sub-ns asymmetric delay attack

Yang Li[1], Jinlong Hu[1], Li Ma[1], Wei Huang[1], Shuai Zhang[1], Yujie Luo[1], Chuang Zhou[1], Chenlin Zhang[1], Heng Wang[1], Yan Pan[1], Yun Shao[1], Yichen Zhang[2], Xing Chen[2], Ziyang Chen[3], Song Yu[2], Hong Guo[3†], and Bingjie Xu[1*]

[1] *Science and Technology on Communication Security Laboratory, Institute of Southwestern Communication, Chengdu 610041, China*
[2] *State Key Laboratory of Information Photonics and Optical Communications, Beijing University of Posts and Telecommunications, Beijing 100876, China*
[3] *State Key Laboratory of Advanced Optical Communication Systems and Networks, Department of Electronics, and Center for Quantum Information Technology, Peking University, Beijing 100871, China*

*†hongguo@pku.edu.cn*
*\*xbjpku@pku.edu.cn*

**Abstract:** Two-way fiber-optic time transfer is a promising precise time synchronization technique with sub-nanosecond accuracy. However, asymmetric delay attack is a serious threat which cannot be prevent by any encryption method. In this paper, a dynamic model based scheme is proposed to defense the sub-nanosecond asymmetric delay attack. A threshold is set according to the estimated time difference by a two-state clock model where the fixed frequency difference is excluded from the time difference to detect the asymmetric delay attack which is smaller than the time difference induced by the fixed frequency difference. Theoretical simulation and experimental demonstration are implemented to prove the feasibility of the scheme. A two-way fiber-optic time transfer system with time stability with 24.5ps, 3.98ps, and 2.95ps at 1s, 10s, and 100s averaging time is shown under sub-ns asymmetric time delay attack experimentally. The proposed method provides a promising secure sub-ns precise time synchronization technique against asymmetric delay attack.

## 1. Introduction

Precise time synchronization has become increasingly important for transportation [1], smart grid [2], contemporary space geodesy [3], high-resolution radio-astronomy [4], and modern particle physics [5]. Among various kinds of time transfer techniques, the two-way time transfer technique is a promising one which transmits time signal symmetrically in both directions to cancel the time jitter for one-way transfer [6].

Two-way satellite time transfer, has achieved nanosecond accuracy [7] and a time stability of 200ps [8]. Due to widely installed optical fiber infrastructure and high short-term stability, two-way fiber-optic time transfer (TWFTT) [3, 9] has attracted attention in recent years with the advantages of low cost. Based on directly measuring arrival times of pulses, a time accuracy as low as tens of picoseconds has been reported [10-13].

For TWFTT technology, the high precision is based on the assumption that the propagation delays are symmetric in the two directions for the synchronization channel [9]. However, the adversary can introduce unknown asymmetric time delays to break the assumption which cannot be prevented by encryption methods. Similar asymmetric delay attack and solutions has been studied for time synchronization protocols like NTP and PTP [14-19]. In [14], the delay attacks for NTP and PTP are studied, and it proposes to measure the propagation delays during initialization of clock synchronization and to monitor the propagation delays during the normal operation of the time synchronization protocol. However, no quantitative analysis on the countermeasure is provided. In [15], requirements for secure two-way protocol, such as IEEE 1588, are proposed. However, the influence of fixed frequency difference between the two parties is not excluded from the adversary detection. In [16], a game theoretic analysis of delay attacks is studied, and a multiple paths strategy is proposed to mitigate delay attack. However, this method needs multiple paths between the master and slave clocks. In [17], a multiple master clocks strategy is proposed to mitigate delay attack. However, this method needs multiple master clocks. In [18], a detection and mitigation model for delay attack is studied both in theory and experimentally. However, only 100 ns level synchronization error is achieved by the method. In [19], model-based and data-based methods are proposed as countermeasure for time attack. However, it does not provide real-time detection of the attack. In all, due to the high precision for TWFTT which can provide sub-nanosecond time synchronization, sub-nanosecond delay attack can still influence the performance seriously. However, a real-time detection and mitigation method for sub-nanosecond asymmetric delay attack is still an open question for TWFTT system.

In this paper, we investigate the method to protect TWFTT from sub-nanosecond asymmetric time delay attack. By analyzing the principle of asymmetric time delay attack, we proposal a defense scheme based on the clock dynamics. Then theoretical simulation and experimental demonstration are implemented to prove the feasibility of this scheme. A TWFTT system with time stability with 26.4ps, 6.82ps, and 3.58ps under sub-nanosecond equal interval asymmetric time delay attack and with 24.5ps, 3.98ps, and 2.95ps under sub-nanosecond random interval asymmetric time delay attack at 1s, 10s, and 100s averaging time is shown. The experimental results show that, almost all the asymmetric delay attack with sub-nanosecond delay can be detected and mitigated by this scheme no matter the attack happens in equal interval or randomly. It provides an efficient method to prevent the asymmetric time delay attack for TWFTT.

## 2. Schematic description

### 2.1 Two-Way Fiber-Optic Time Transfer Scheme

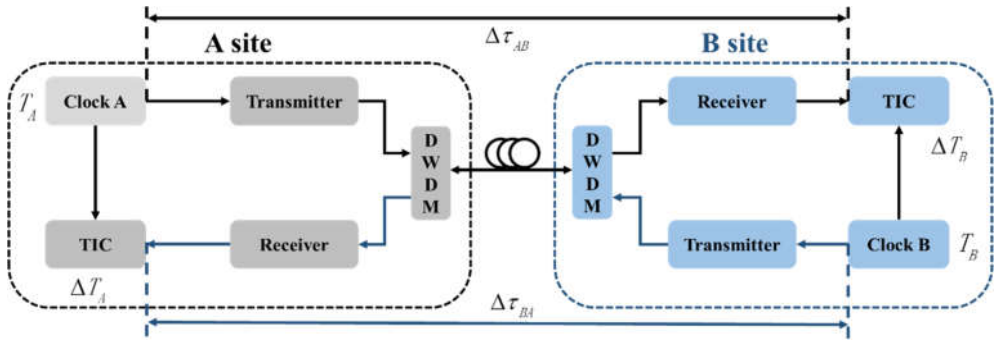

Fig. 1. Schematic diagram of TWFTT system. The designation of equipment and measurement quantities are detailed in section 2.1. DWDM: Dense Wavelength Division Multiplexing, TIC: Time Interval Counter.

We consider a general scheme of the TWFTT system (see in Fig. 1). It consists of two parties, A and B, interconnected by an optical fiber channel. At each of the parties, the time scales (1PPS) are transmitted to each other through a bidirectional optical fiber link. The 1PPS signals are detected by the receiver. The time difference between the received 1PPS and the sent 1PPS are measured by the time interval counter (TIC) at each part, named $\Delta T_A$ and $\Delta T_B$. So,

$$\Delta T_A = (T_B + \Delta \tau_{BA}) - T_A,\tag{1}$$

$$\Delta T_B = (T_A + \Delta \tau_{AB}) - T_B, \qquad (2)$$

where $\Delta \tau_{BA}$ is the propagation time from B to A, and $\Delta \tau_{BA}$ in the other direction.

According to Eq. (1) and Eq. (2), the time offset between A and B is derived by

$$\Delta T = T_A - T_B = \frac{1}{2}(\Delta T_B - \Delta T_A) + \frac{1}{2}(\Delta \tau_{BA} - \Delta \tau_{AB}). \qquad (3)$$

Assuming a symmetrical propagation delay, $\Delta \tau_{BA} = \Delta \tau_{AB}$, the time offset between A and B is given by

$$\Delta T = \frac{1}{2}(\Delta T_B - \Delta T_A). \qquad (4)$$

For TWFTT system, one party is called the remote clock and the other is called the local clock. The measured time offset is used by local clock to correct its clock to synchronizing with the remote clock. In this paper, B is used as the local clock, and A is used as the remote clock.

*2.2 Asymmetric delay attack*

Asymmetric delay in the channel will lead to synchronization errors in TWFTT system. An adversary can disturb the performance of TWFTT system by introducing malicious time delays in one of the transmitting channels. The attack is called asymmetric delay attack.

The impact of the asymmetric delay attack is analyzed quantitatively. Specifically, if the adversary delays the 1PPS from B to A by $\Delta \tau_{attack}$, the actual time offset between A and B is given by

$$\Delta T_{actual} = \frac{1}{2}(\Delta T_B - \Delta T_A) + \frac{1}{2}\Delta \tau_{attack}. \qquad (5)$$

Comparing Eq.(4) and Eq.(5), if the parties do not find the adversary, the adversary introduces time synchronization error with $\Delta \tau_{error} = \frac{1}{2}\Delta \tau_{attack}$.

*2.3 Countermeasure*

In order to detect the asymmetric delay attack, we build a clock dynamic model for the time offset between A and B. An adversary detector function with the measured time offset and the estimated time offset from the dynamic model as variables is built. By setting a security threshold, if the value of the detector exceeds the threshold, a potential attack is detected, and a special time offset correction scheme is chosen. And if the value of the detector does not exceed the threshold, a normal time offset correction scheme is chosen.

A clock can be considered as an oscillator and a counter. In this paper, frequency drift and aging are not considered, so, a two-state clock model is employed [20]. Equations describing the clock dynamics are

$$\begin{cases} d\theta(t) = \gamma(t)dt + d\omega_\theta(t) \\ d\gamma(t) = d\omega_\gamma(t) \end{cases}, \qquad (6)$$

where $\theta(t) = t_{local} - t_{remote}$ and $\gamma(t) = (f_{local} - f_{remote})/f_{remote}$ are time offset and skew between the local clock and the remote clock, $\omega_\theta(t)$ and $\omega_\gamma(t)$ relate to random-walk phase noise and random-walk frequency noise respectively, which are independent one-dimensional zero-mean Wiener processes with variances equal to $\sigma_\theta^2$ and $\sigma_\gamma^2$ respectively.

For TWFTT system, the local clock is updated periodically, and time offset correction, $u_\theta(t_n)$, is applied to the local clock to synchronizing with the remote clock at the *n*th synchronization instant $t_n = n \cdot \tau$. So, Eq. (6) can be rewritten as difference equations [21]:

$$\begin{cases} \theta(t_n) = \theta(t_{n-1}) + u_\theta(t_{n-1}) + \gamma(t_{n-1}) \cdot \tau + \omega_\theta(t_n) \\ \gamma(t_n) = \gamma(t_{n-1}) + \omega_\gamma(t_n) \end{cases}. \quad (7)$$

For TWFTT system, the measured time offset, which is rewrite as $\theta_M(t_n) = \Delta T(t_n)$, can be used to correct the local clock. Correction strategy influences the performance of the TWFTT system and adversary detection effect. In this paper, we focus on the direct correction strategy.

For the direct correction strategy, the measured time offset is used to correct the local clock directly, $u_\theta(t_n) = \theta_M(t_n)$. In order to detect the delay attack on the channel, the frequency difference between the local clock and remote is supposed to be fixed. The frequency difference can bring fixed time offset, which needs to be removed to construct the detector. In this paper, we propose an algorithm without estimating the asymmetric delay value brought by the attack.

---

**Algorithm: Asymmetric Time Delay Attack Detection**

1. Calculate measured time offset, $\theta_M(t_n) = \Delta T(t_n) = \frac{1}{2}(\Delta T_B(t_n) - \Delta T_A(t_n))$.

2. Calculate the estimated frequency difference.
   If no attack detected at $t_{n-1}$, then the measured frequency difference at $t_n$ is given by $\gamma_M(t_n) = (\theta_M(t_n) - \theta_M(t_{n-1}) + u_\theta(t_{n-1}))/\tau$, and the estimated frequency difference is given by $\hat{\gamma}_E(t_n) = \gamma_M(t_n)$.
   Else, the estimated frequency difference is given by $\hat{\gamma}_E(t_n) = \gamma_{best}(t_{n-1})$.

3. Calculate the fixed time offset induced by frequency difference, $offsetF(t_n) = \gamma_{best}(t_{n-1}) \cdot \tau$.

4. Calculate the attack index, $I_{attack} = |\theta_M(t_n) - offsetF(t_n)|$.

5. Make a judgment whether an attack happens at $t_n$.
   If $I_{attack} > I_{threhold}$, then $u_\theta(t_n) = offsetF(t_n)$, and $\gamma_{best}(t_n) = w * \gamma_{best}(t_{n-1}) + (1-w) * \gamma_{best}(t_{n-2})$, $0 \le w \le 1$.
   Else, $u_\theta(t_n) = \theta_M(t_n)$, $\gamma_{best}(t_n) = w * \gamma_M(t_n) + (1-w) * \gamma_{best}(t_n)$, $0 \le w \le 1$.

---

Table 1. Simulation parameter

| Parameter | Value | Description |
|---|---|---|
| $\sigma_m$ | 25ps | Standard deviation of measurement noise |
| $\sigma_d$ | 10ps | Standard deviation of transmission noise |
| $\sigma_\theta$ | 10ps | Standard deviation of random-walk phase noise |
| $\sigma_\gamma$ | 1ps/s | Standard deviation of random-walk frequency noise |

*2.4 Metric*

On the one hand, in order to analysis the effect of the attack detection algorithm quantitatively, two performance metrics are introduced which are precision and recall. Precision is defined as the number of detected actual attack events over the total number of detected attack events, while recall is defined as the number of detected actual attack events over the total number of actual attack events.

On the other hand, in order to analysis the influence of the attack detection on the performance of the time synchronization system, time deviation error variance (TDEV) and maximum time interval error (MTIE) are introduced [22].

$$\text{TDEV}(\tau = n*\tau_0) = \sqrt{\frac{1}{6n^2(N-3n+1)} \sum_{j=0}^{N-3n} [\sum_{i=j}^{n+j-1} (x_{i+2n} - 2x_{i+n} + x_i)]^2} \quad (8)$$

$$\text{MTIE}(\tau = n*\tau_0) = \max_{i=0}^{N-n-1} \{\max_{k=1}^{k=n}[|x(i+k) - x(i)|]\} \quad (9)$$

### 3. Theoretical Simulation

In this section, theoretical simulation is implemented to prove the feasibility of the attack detection strategy proposed in this paper.

Firstly, the performance of TWFTT system under no attack with direct correction strategy and attack detection strategy is compared by simulation. In the simulation, two independent one-dimensional zero-mean Wiener processes with variances $\sigma_m^2$ and $\sigma_d^2$ are introduced for the measurement noise and transmission noise. Without loss of generality, all the values of the noises in the simulation are chosen as in Table 1. As shown in Fig. 2, the time synchronization error (time offset) is just around zero for both cases where the fluctuation is caused by measurement noise, transmission noise and random-walk noise.

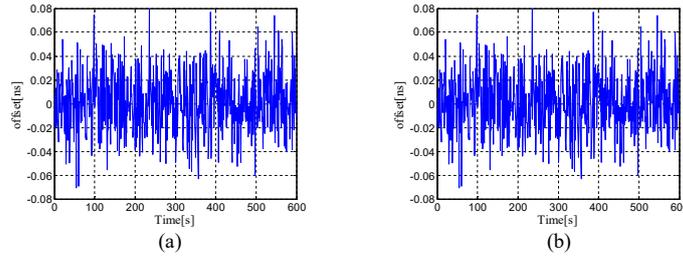

(a)　　　　　　　　　　　　　(b)

Fig. 2. Simulation of TWTFF's time difference without attack. (a) direct correction strategy; (b) attack detection strategy.

In order to evaluate the influence of the attack detection algorithm on the performance of time synchronization quantitatively, we studied the TDEV and MTIE for both cases. As shown in Fig. 3, time stability with metrics TDEV and MTIE are almost the same for the two cases. The results show that the attack detection algorithm does not deteriorate the performance of the time synchronization if no attack exists.

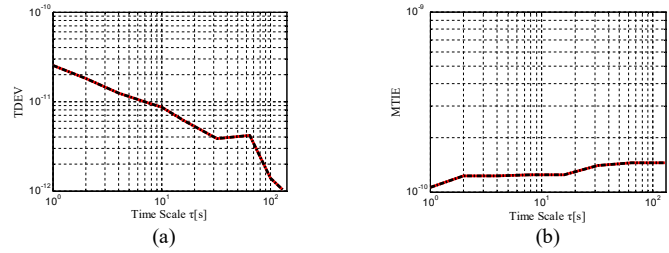

(a)　　　　　　　　　　　　　(b)

Fig. 3. Simulation of TWTFF's TDEV and MTIE without attack. (a) TDEV; (b) MTIE.

(solid line(red): attack detection strategy; dash line(black): direct correction strategy).

Secondly, the influences of delay attack on the time synchronization are studied. In the simulation, an delay attack happens once every 50s. Two cases are compared, where direct correction strategy is adapted for the first case and attack detection strategy is adapted for the second case.

For the case of direct correction strategy, since no attack detection is adapted, the time delay attack brings in the synchronization errors (see Fig. 4(a, b, c)). As shown in Eq. (4) and Eq. (5), the offset correction algorithm does not recognize the time delay introduced by the adversary, and a time synchronization error with $\frac{1}{2}\Delta\tau_{attack}$ is concluded in the offset correction calculation. In the simulation, three case are studied, where the theoretical time synchronization error $\Delta\tau_{error}$ is 1ns, 0.5ns, 0.2ns. As shown in Fig.4(a, b, c), the actual time error with values around 1ns, 0.5ns, 0.2ns is brought every 50s. The fluctuation is caused by the noise, such as measurement noise, transmission noise and random-walk noise. The simulation result matches with the theoretical analysis.

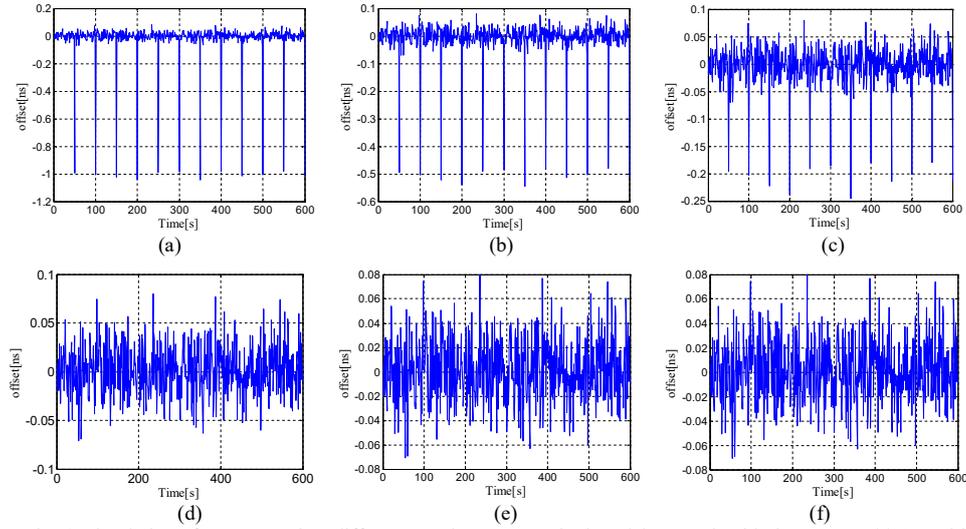

Fig. 4. Simulation of TWTFF's time difference under asymmetric time delay attack with time errors: (a) 1ns with direct correction strategy; (b) 0.5ns with direct correction strategy; (c) 0. 2ns with direct correction strategy; (d) 1ns with attack detection strategy; (e) 0.5ns with attack detection strategy; (f) 0.2ns with attack detection strategy.

For the case of attack detection strategy, all the actual attack events are detected by the algorithm, and no event which is not attack event is recognized as attack event. So, precision and recall for the simulation are both 100%. From Fig. 4(d, e, f), we can see that the actual time offset is around zero, and the influence of the delay attack is eliminated by the algorithm.

In order to evaluate the influence quantitatively, TDEV and MTIE curves are drawn (see Fig. 5). By comparing the direct strategy without attack and with attack, the results show that the delay attack brings serious influence on the performance of the time synchronization. By comparing attack detection strategy without attack and with attack, the results show that the influence of the delay attack can be effectively eliminated by the attack detection strategy.

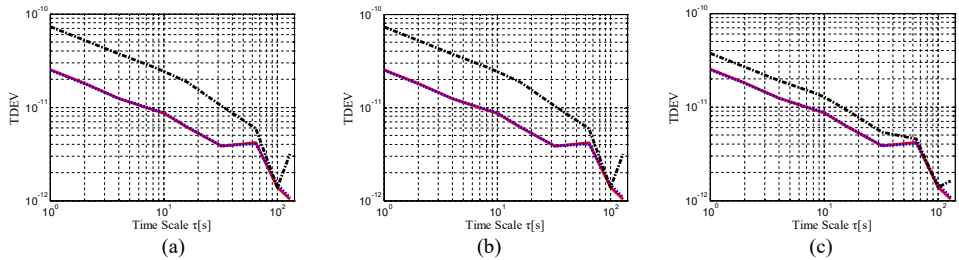

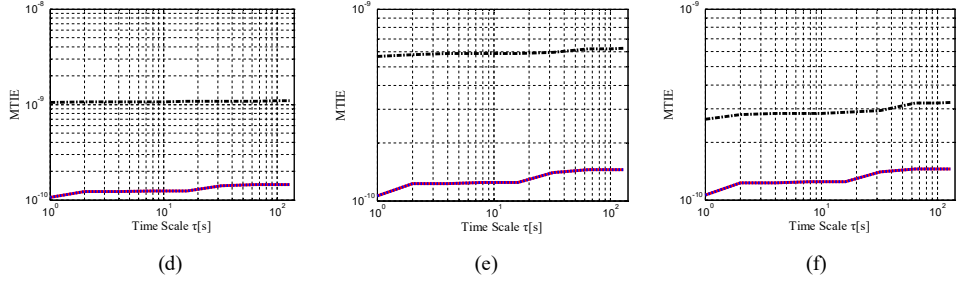

Fig. 5. Simulation of TWTFF's TDEV and MTIE under asymmetric time delay attack with time errors: (a) TDEV under 1ns attack; (b) TDEV under 0.5ns attack; (c) TDEV under 0.2ns attack; (d) MTIE under 1ns attack; (e) MTIE under 0.5ns attack; (f) MTIE under 0.2ns attack. (solid line(red): direct correction strategy without attack; dash line(blue): attack detection strategy with attack; dot line(black): direct correction strategy with attack).

As shown in Table 2, TDEVs and MTIEs at average time 1s, 10s, and 100s are compared. By comparing cases of attack detection strategy without attack, and with 1ns, 0.5 ns and 0.2ns attack respectively, it shows that the proposed attack detection algorithm can distinguish effectively the attack events and the normal events, and all the TDEVs and MTIEs at different average time are almost the same. For TDEV@1s, TDEV@10s, TDEV@100s, MTIE@1s, MTIE@10s, and MTIE@100s, all the cases are around 25ps, 8.7ps, 1.3ps, 106ps, 125ps, and 145ps, respectively. By comparing cases of direct correction strategy and cases of attack detection strategy with 1ns, 0.5ns and 0.2ns attack, it shows that the asymmetric time delay attack brings serious influence on the performance of the time synchronization. Counter-intuitively, TDEV@100s is seemed to not be influenced by the attack. It is caused by the definition of TDEV. According to Eq. (8), when $\tau = 100$, $x_{i+2n} - 2x_{i+n} + x_i$ equals to $x_{i+200} - 2x_{i+100} + x_i$. Since the interval of attack event in the simulation is 50s, when $i$ is an integral multiple of 50, the same time error is induced by the attack for $x_{i+200}$, $x_{i+100}$ and $x_i$, so the effects are counteracted, and when $i$ is not an integral multiple of 50, no time error is induced by the attack for $x_{i+200}$, $x_{i+100}$ and $x_i$. So, the TDEV curve of the direct correction strategy and the TDEV curve of no attack case meet when $\tau = 100$ as shown in Fig. 5, and the TDEVs@100s are almost the same for the cases of attack detection strategy and the cases of direct correction strategy as shown in Table 2.

Table 2. Performance metric under asymmetric time delay attack

| | Recall | Precision | TDEV @1s | TDEV @10s | TDEV @100s | MTIE @1s | MTIE @10s | MTIE @100s |
|---|---|---|---|---|---|---|---|---|
| Attack Detection strategy without attack | / | / | 2.5174 E-11 | 8.6769 E-12 | 1.3905 E-12 | 1.0647 E-10 | 1.2489 E-10 | 1.4537 E-10 |
| Attack Detection strategy with 1ns attack | 100% | 100% | 2.5012 E-11 | 8.7183 E-12 | 1.4889 E-12 | 1.0647 E-10 | 1.2489 E-10 | 1.4537 E-10 |
| Direct correction strategy with 1ns attack | / | / | 1.3820 E-10 | 4.4448 E-10 | 4.0823 E-12 | 1.0462 E-09 | 1.0688 E-09 | 1.0962 E-09 |

| | | | | | | | |
|---|---|---|---|---|---|---|---|
| Attack Detection strategy with 0.5ns attack | 100% | 100% | 2.5012 E-11 | 8.7183 E-12 | 1.4889 E-12 | 1.0647 E-10 | 1.2489 E-10 | 1.4537 E-10 |
| Direct correction strategy with 0.5ns attack | / | / | 7.2929 E-11 | 2.4092 E-11 | 1.3905 E-12 | 5.6548 E-10 | 5.8463 E-10 | 2.6192 E-10 |
| Attack Detection strategy with 0.2ns attack | 100% | 100% | 2.5012 E-11 | 8.7183 E-12 | 1.4889 E-12 | 1.0647 E-10 | 1.2489 E-10 | 1.4537 E-10 |
| Direct correction strategy with 0.2ns attack | / | / | 3.7248 E-11 | 1.2712 E-11 | 1.3904 E-12 | 1.6549 E-10 | 2.8467 E-10 | 3.2196 E-10 |

## 4. Experimental Demonstration

In this section, a TWFTT system is set up in the laboratory and experimental demonstration are implemented to prove the feasibility of the attack detection strategy proposed in this paper.

The experimental setup of TWFTT system with adversary simulator is shown in Fig. 6. On the remote/local site, the digital delay generator (DDG, SRS DG645) generates 1PPS electric signal. 1PPS from one output port of DDG drives an electro-optic modulator (EOM, AX-0S5-10-PFA-PFA-UL) to modulate the CW laser to generate 1PPS optic signal. The same 1PPS from another output port of DDG is sent to the start trigger port of a time interval counter (TIC). The 1PPS optic signal is coupled to channel 35 of DWDM, and transmitted to the local site through the fiber channel. The photo detector (PD) on the local/remote site detects the 1PPS optical signal and the generated electric signal is sent to the stop trigger port of the TIC on the local/remote site. The time difference recorded by TIC on remote site is sent to local site. By Eq. (4), the measured time offset is calculated on local site. According to the adversary detection strategy and correction strategy, the delay correction value is calculated on the computer of the local site and sent to DDG on the local site to modify the time delay. In order to evaluate the strategy, an extra TIC is added to measure the actual time errors between remote clock and local clock.

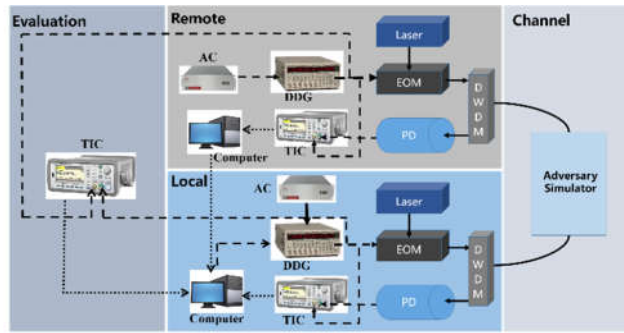

Fig. 6. Experimental setup of Two-Way Fiber-optic Time Transfer system with adversary simulator. AC: Atomic clock, TIC: Time Interval Counter, DDG: Digital Delay Generator, EOM: Electro-Optic Modulator, PD: Photo Detector, DWDM: Dense Wavelength Division Multiplexing.

An adversary simulator is installed in the fiber channel, which can simulate the asymmetry delay attack launched by the adversary. Similar to [23], the adversary simulator consists two 1x4 optical switches. When the two optical switches are set to path 1, no asymmetry delay is added to the channel. When the two optical switches are set to path 2, 0.296 ns asymmetry delay is added to the channel. When the two optical switches are set to path 3, 0.83 ns asymmetry delay is added to the channel. When the two optical switches are set to path 4, 1.25 ns asymmetry delay is added to the channel.

Before studying the influence of the attack on the TWFTT system, we first compare the attack detection strategy and direct correction strategy without attack. For each strategy, 600s data of evaluation TIC are recorded, and the TDEV and MTIE are calculated, as shown in Fig. 7 and Fig. 8. Different from the simulation case, the values of TDEV and MTIE between the attack detection strategy and direction correction strategy are not the same. Because in the simulation, the measurement noise, transmission noise and process noise are the same for the two strategies, and in the experiment, these noises are different for the two strategies. However, although the values of TDEV and MITE are the exactly the same for the two strategies, the values are very close. Many experiments are done to confirm difference between the two strategies is induced by the random noises.

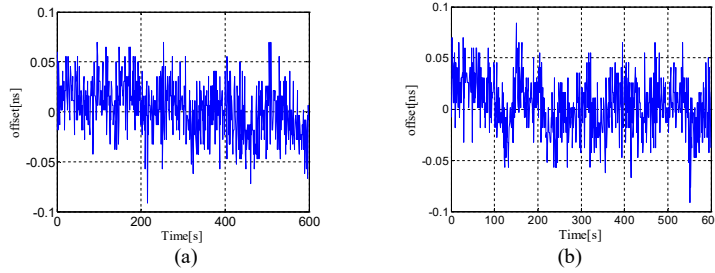

Fig. 7. TWTFF's time difference without attack. (a) direct correction strategy; (b) attack detection strategy.

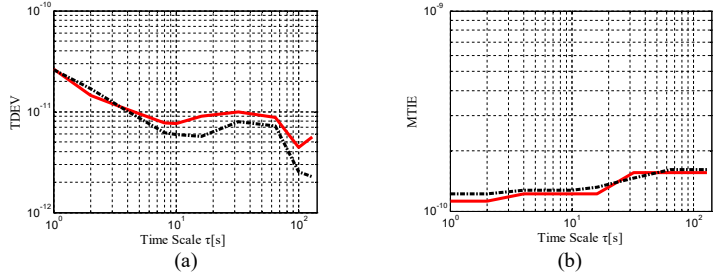

Fig. 8. TWTFF's TDEV and MTIE without attack (solid line(red): attack detection strategy; dash line(black): direct correction strategy). (a) TDEV; (b) MTIE

As shown in Table 3, TDEVs and MTIEs at average time 1s, 10s, and 100s are compared for attack detection strategy and direct correction strategy without attack. It shows that TDEV@1s, TDEV@10s, TDEV@100s, MTIE @1s, MTIE@10s and MTIE@100s are around 26.5ps, 7.6ps, 4.4ps, 112.3ps, 122.1ps, and 156.3ps for attack detection strategy and 26.1ps, 5.9ps, 2.5ps, 122.1ps, 127.0ps, and 161.1ps for direct correction strategy. The difference is caused by the different noises in the experiments.

**Table 3. Performance metric without attack**

|  | TDEV @1s | TDEV @10s | TDEV @100s | MTIE @1s | MTIE @10s | MTIE @100s |
|---|---|---|---|---|---|---|
| Attack Detection strategy without attack | 2.6534 E-11 | 7.6253 E-12 | 4.4366 E-12 | 1.1230 E-10 | 1.2207 E-10 | 1.5625 E-10 |
| Direct correction strategy without attack | 2.6096 E-11 | 5.9026 E-12 | 2.5161 E-12 | 1.2207 E-10 | 1.2695 E-10 | 1.6113 E-10 |

In this paper, two kinds of asymmetry delay attacks are studied, the equal interval attack and random interval attack.

### 4.1 Equal interval attack

For equal interval attack, the adversary launched asymmetry delay attack, once at set intervals. Without loss of generality, we set 50s as the interval. Three kinds of asymmetry delay attack with 0.296ns, 0.83ns and 1.25ns delay respectively are studied.

As shown in Fig.9(a, b, c), when no attack detection strategy is applied, the TWFTT system can not recognize the attacks, and large synchronization errors are brought by the asymmetry delay attack. In order to evaluate the influence quantitatively, TDEV and MTIE curves are drawn, as shown in Fig. 10. The results show that the equal interval attack brings serious influence on the performance of the time synchronization, and the influence of the delay attack is eliminated by the attack detection strategy.

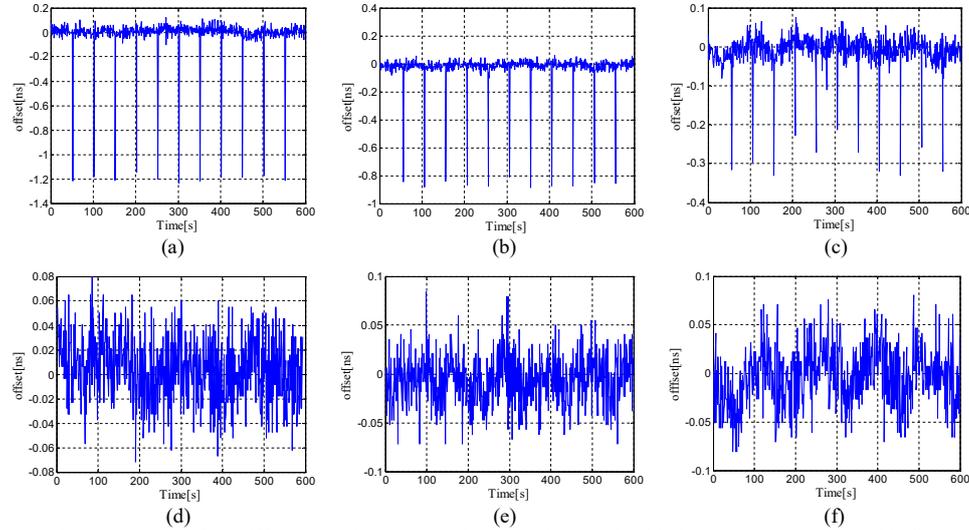

Fig. 9. TWFF's time difference under asymmetric time delay attack with time errors: (a) 1.25ns with direct correction strategy; (b) 0.83ns with direct correction strategy; (c) 0. 296ns with direct correction strategy; (d) 1.25ns with attack detection strategy; (e) 0.83ns with attack detection strategy; (f) 0. 296ns with attack detection strategy.

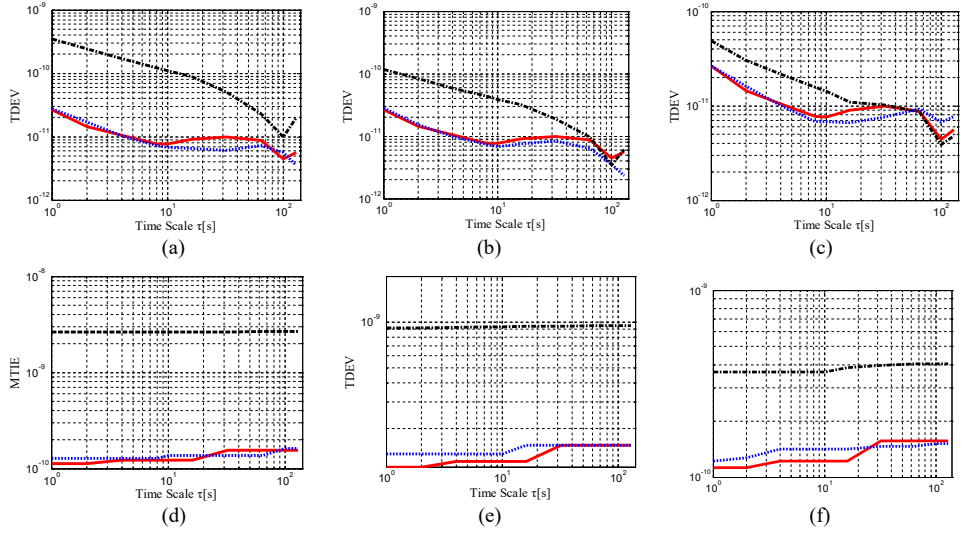

Fig. 10. TWTFF's TDEV and MTIE under asymmetric time delay attack with time errors: (a) TDEV under 1.25ns attack; (b) TDEV under 0.83ns attack; (c) TDEV under 0.296ns attack; (d) MTIE under 1.25ns attack; (e) MTIE under 0.83ns attack; (f) MTIE under 0.296ns attack. (solid line(red): direct correction strategy without attack; dash line(blue): attack detection strategy with attack; dot line(black): direct correction strategy with attack).

As shown in Table 4, TDEVs and MTIEs at average time 1s, 10s, and 100s are compared. By comparing cases of attack detection strategy without attack, with 1.25ns attack, with 0.83 ns attack and with 0.296ns attack, it shows that the attack detection algorithm proposed in this paper can distinguish the attack events and the normal events. The differences of TDEVs and MTIE are caused by the difference of the noises. The smallest TDEVs at average time 1s, 10s and 100s are 26.4ps, 6.82ps, and 3.58ps, and the smallest MTIEs at average time 1s, 10s and 100s are 122.1.ps, 136.7ps and 151.4ps, under nanosecond and sub-nanosecond equal interval attacks. By comparing cases of direct correction strategy and cases of attack detection strategy with 1.25ns attack, 0.83ns attack and 0.296ns attack, it shows that the asymmetric time delay attack brings serious influence on the performance of the time synchronization. Counter-intuitively, TDEV@100s is seemed to be not influenced by the attack. It is caused by the definition of TDEV. According to Eq. (8), when $\tau = 100$, $x_{i+2n} - 2x_{i+n} + x_i$ equals to $x_{i+200} - 2x_{i+100} + x_i$. Since the interval of attack event in the simulation is 50s, when i is an integral multiple of 50, the same time error is induced by the attack for $x_{i+200}$, $x_{i+100}$ and $x_i$, so the effects are counteracted, and when $i$ is not an integral multiple of 50, no time error is induced by the attack for $x_{i+200}$, $x_{i+100}$ and $x_i$. So, the TDEV value of the direct correction strategy approach the TDEV value without attack as shown in Fig. 10.

Table 4. Performance metric under equal interval attack

| | Recall | Precision | TDEV @1s | TDEV @10s | TDEV @100s | MTIE @1s | MTIE @10s | MTIE @100s |
|---|---|---|---|---|---|---|---|---|
| Attack Detection strategy without attack | / | / | 2.6534 E-11 | 7.6253 E-12 | 4.4366 E-12 | 1.1230 E-10 | 1.2207 E-10 | 1.5625 E-10 |
| Attack Detection strategy with 1.25 ns attack | 100% | 100% | 2.7675 E-11 | 6.8526 E-12 | 5.6365 E-12 | 1.2695 E-10 | 1.3672 E-10 | 1.6113 E-10 |
| Direct correction strategy with 1.25ns attack | / | / | 1.8109 E-10 | 5.9854 E-11 | 8.0417 E-12 | 1.3818 E-09 | 1.4063 E-09 | 1.4795 E-09 |
| Attack Detection strategy with 0.83ns attack | 100% | 100% | 2.7934 E-11 | 6.8546 E-12 | 3.5781 E-12 | 1.3672 E-10 | 1.3672 E-10 | 1.5625 E-10 |
| Direct correction strategy with 0.83ns attack | / | / | 1.1770 E-10 | 3.8755 E-11 | 3.5045 E-12 | 9.1797 E-10 | 9.3262 E-10 | 9.4727 E-10 |
| Attack Detection strategy with 0.296ns attack | 100% | 100% | 2.6409 E-11 | 6.8165 E-12 | 6.7124 E-12 | 1.4160 E-10 | 1.4160 E-10 | 1.6602 E-10 |
| Direct correction strategy with 0.296ns attack | / | / | 4.8687 E-11 | 1.4371 E-11 | 3.8927 E-12 | 3.6621 E-10 | 3.6621 E-10 | 4.0527 E-10 |

### 4.2 Random interval attack

For random interval attack, the adversary launched asymmetry delay attack randomly. So, attack can happened consecutively. The probability of the attack is a key parameter. Without loss of generality, three kinds of random interval attack are studied. The first kind is 0.83ns delay random interval attack with $p_{no}=0.8$, $p_{0.83ns}=0.2$. The second kind is 0.296ns delay random interval attack with $p_{no}=0.8$, $p_{0.296ns}=0.2$. The third kind is mixed 0.83ns and 0.296ns delay random interval attack with $p_{no}=0.7$, $p_{0.83ns}=0.15$, $p_{0.296ns}=0.15$.

As shown in Fig.11, when no attack detection strategy is applied, the TWFTT system can not recognize the attacks, and large synchronization errors are brought by the asymmetry delay attack. In order to evaluate the influence quantitatively, TDEV and MTIE curves are drawn, as shown in Fig. 12. The results show that the random interval attack brings serious influence on the performance of the time synchronization, and the influence of the delay attack is eliminated by the attack detection strategy.

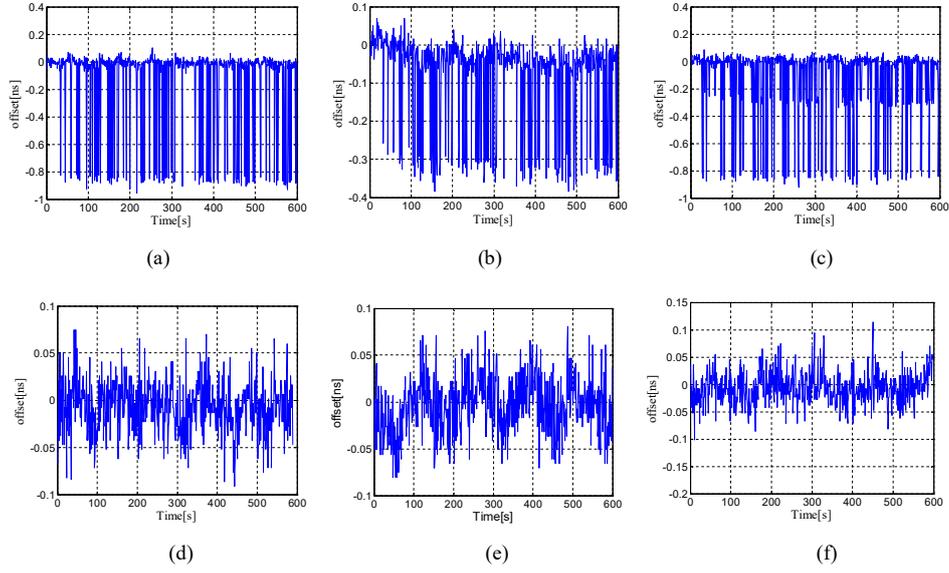

Fig. 11. TWTFF's time difference under asymmetric time delay attack with time errors: (a) $p_{no}=0.8$, $p_{0.83ns}=0.2$ with direct correction strategy; (b) $p_{no}=0.8$, $p_{0.296ns}=0.2$ with direct correction strategy; (c) $p_{no}=0.7$, $p_{0.83ns}=0.15$, $p_{0.296ns}=0.15$ with direct correction strategy; (d) $p_{no}=0.8$, $p_{0.83ns}=0.2$ with attack detection strategy; (e) $p_{no}=0.8$, $p_{0.296ns}=0.2$ with attack detection strategy; (f) $p_{no}=0.7$, $p_{0.83ns}=0.15$, $p_{0.296ns}=0.15$ with attack detection strategy.

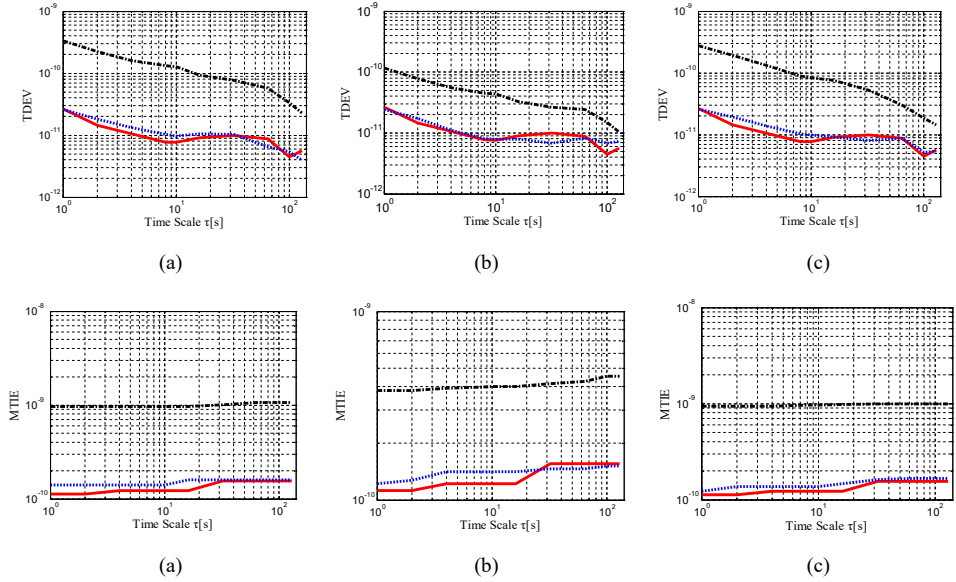

Fig. 12. TWTFF's TDEV and MTIE under asymmetric time delay attack with time errors: (a) TDEV under $p_{no}=0.8$, $p_{0.83ns}=0.2$; (b) TDEV under $p_{no}=0.8$, $p_{0.296ns}=0.2$; (c) TDEV under $p_{no}=0.7$, $p_{0.83ns}=0.15$, $p_{0.296ns}=0.15$; (d) MTIE under $p_{no}=0.8$, $p_{0.83ns}=0.2$; (e) MTIE under $p_{no}=0.8$, $p_{0.296ns}=0.2$; (f) MTIE under $p_{no}=0.7$, $p_{0.83ns}=0.15$, $p_{0.296ns}=0.15$. (solid line(red): direct correction strategy without attack; dash line(blue): attack detection strategy with attack; dot line(black): direct correction strategy with attack).

As shown in Table 5, TDEVs and MTIEs at average time 1s, 10s, and 100s are compared. By comparing cases of attack detection strategy without attack, with 0.83ns attack, with 0.296 ns attack and with 0.2ns attack, and 0.83ns & 0.296ns mixed attack, it shows that the attack detection algorithm proposed in this paper can distinguish the attack events and the normal events. The differences of TDEVs and MTIE are caused by the difference of the noises. The smallest TDEVs at average time 1s, 10s and 100s are 24.5ps, 3.98ps, and 2.95ps, and the smallest MTIEs at average time 1s, 10s and 100s are 122.1.ps, 136.7ps and 151.4ps, under sub-nanosecond random interval attacks. By comparing cases of attack detection strategy with cases of direct correction strategy, it shows that, different from the equal interval attack, the TDEVs of the direct correction strategy does not approach the TDEV value without attack when $\tau=100$ since the interval of the attack event is random.

Table 5. Performance metric under random interval attack

| | Recall | Precision | TDEV @1s | TDEV @10s | TDEV @100s | MTIE @1s | MTIE @10s | MTIE @100s |
|---|---|---|---|---|---|---|---|---|
| Attack Detection strategy without attack | / | / | 2.6534 E-11 | 7.6253 E-12 | 4.4366 E-12 | 1.1230 E-10 | 1.2207 E-10 | 1.5625 E-10 |
| Attack Detection strategy with 0.83ns attack | 100% | 100% | 2.8298 E-11 | 3.9795 E-12 | 2.9450 E-12 | 1.2695 E-10 | 1.5905 E-10 | 1.6113 E-10 |
| Direct correction strategy with 0.83ns attack | / | / | 3.3061 E-10 | 1.2786 E-10 | 3.3645 E-11 | 9.6680 E-10 | 9.6680 E-10 | 1.0596 E-09 |
| Attack Detection strategy with 0.296ns attack | 100% | 100% | 2.4511 E-11 | 7.9209 E-12 | 6.7849 E-12 | 1.2207 E-10 | 1.4160 E-10 | 1.5137 E-10 |
| Direct correction strategy with 0.296ns attack | / | / | 1.1571 E-10 | 4.3600 E-11 | 1.4957 E-11 | 3.8086 E-10 | 4.0039 E-10 | 4.5410 E-10 |
| Attack Detection strategy with 0.83ns & 0.296ns mixed attack | 100% | 100% | 2.5856 E-11 | 9.9418 E-12 | 5.1802 E-12 | 1.2207 E-10 | 1.3672 E-10 | 1.6602 E-10 |
| Direct correction strategy with 0.83ns & | / | / | 4.9180 E-10 | 1.8902 E-10 | 4.2464 E-11 | 1.3818 E-09 | 1.3818 E-09 | 1.4014 E-09 |

| | |
|---|---|
| | 0.296ns mixed attack |

## 5. Discussion and conclusion

TWFTT system is a kind of technology to provide sub-nanosecond time synchronization with low cost. It is based on assumption that propagation delays are symmetric in the two directions for the synchronization channel, so the precision is mainly limited by the residual asymmetrical propagation delays which brings sub-nanosecond uncertainty. However, the adversary can introduce extra asymmetric delays to break the assumption which can influence the performance seriously.

In this paper, we propose a dynamic model based method to protect the TWFTT system from sub-nanosecond asymmetric delay attack. The theoretical simulation shows that the method is effective to protect the TWFTT system. In order to prove the feasibility of the method for the real TWFTT system, the method is tested under to kinds of attacks, equal interval attack and random interval attack. The results show that the effect of the attack is eliminated by the method for the real TWFTT system. In order to measure the performance, TDEV and MTIE are calculated. With this method, a TWFTT system with time stability about 24.5ps, 3.98ps, and 2.95ps at 1s, 10s, and 100s averaging time is shown under sub-ns asymmetric time delay attack. The computation complexity of the proposed method is low and can detect the attack in real-time. So this method can easily be integrated in the TWFTT system to provide secure sub-ns precise time synchronization under asymmetric delay attack. In the future, one the one hand, we experimentally verify the feasibility of the proposed method in the TWFTT system over a 10km fiber link. However, for longer transmission distance, erbium-doped optical fiber amplifier will be integrated in the link as repeater, which distorts the waveform. Part of the distortion is fixed, and the other part is random. So, optimization of the delay attack detection method for longer transmission is an interesting problem. On the other hand, network of TWFTT has attracted much attention recent years. For the network, a node may be an intersection of multiple TWFTT paths. Other path can provide additional information for attack detection. So, systematic delay attack detection method in network is an interesting open question.

## 6. Acknowledgments

The authors would like to thank Dr. Giada Giorgi for fruitful discussions. This work was supported in part by China NSF under Grants 61901425, U19A2076, 61771081, 62101516, 61771439, 61702469, in part by Fundamental Research Funds for the Central Universities under Grant 2019CDXYJSJ0021, in part by Sichuan Youth Science and Technology Foundation under Grants 2019JDJ0060.